\documentclass[a4paper]{spie}  
\usepackage[]{graphicx}

\title{Cold-electron bolometers for future mm and sub-mm sky surveys}

\author{Maria Salatino\supit{a} and Paolo de Bernardis\supit{a} and
Sumedh Mahashabde\supit{b} and Leonid S. Kuzmin\supit{b} and Silvia Masi\supit{a} \skiplinehalf
\supit{a}Physics Department Sapienza University of Rome, piazzale Aldo Moro 5 00185, Rome, Italy; \\
\supit{b}Department of Microtechnology and Nanoscience - MC2
Chalmers University of Technology, SE-412 96, G\"{o}teborg, Sweden}

\authorinfo{Further author information: (Send correspondence to Maria Salatino)\\ Maria Salatino: E-mail: maria.salatino@roma1.infn.it, Telephone: +39 06 49914342}

\begin{document}
  \maketitle

\begin{abstract}
Future sky surveys in the mm/sub-mm range, like the forthcoming balloon-borne missions LSPE, OLIMPO, SPIDER etc., will need
detectors insensitive to cosmic rays (CRs) and with a NEP of the order of $10^{-17} \div 10^{-18}\,$W/sqrt(Hz). The Cold-Electron
Bolometers (CEBs) technology is promising, having the required properties, since the absorber volume is extremely small and the
electron system of the absorber is thermally insulated from the phonon system. We have developed an experimental setup to test the
optical performance and the CRs insensitivity of CEBs, with the target of integrating them in the OLIMPO and LSPE focal planes.
\end{abstract}

\keywords{Bolometers, Cold Electrons, Cosmic Rays, Space
Instrumentation}

\section{INTRODUCTION}\label{sec:intro}
The recent Planck\cite{Planck13} and BICEP2\cite{BICEP2} results have improved our knowledge of the cosmological model and given
the first hint of primordial gravitational waves, respectively. Despite their success, the Planck mission was close to be limited
by glitches produced by cosmic rays\cite{Planck13b}$^{,}$\cite{Catalano14} , while BICEP2 was affected by common limitations of
ground-based observations\cite{BICEP3}. To further improve our present knowledge of the primordial universe and to confirm the tiny
B-modes signal detected by BICEP2, we need independent measurement strategies and completely different technologies. In this
contest future stratospheric\cite{SWIPE12}$^{,}$\cite{Fraisse13} and satellite missions\cite{PRISM14} will be favored, since they
allow to improve both the frequency coverage and the sky coverage, two essential requirements for a reliable measurement of B-modes
and to assess their primordial origin. These missions will require large arrays of ultra sensitive detectors, insensitive to the
ionizing radiation present in space (cosmic rays). A good candidate is the Cold-Electron Bolometer (CEB): a detector based on the
behavior of Superconducting-Insulator-Normal metal junction
\cite{Golubev01}$^{,}$\cite{Kuzmin02}$^{,}$\cite{Kuzminb02}$^{,}$\cite{Tarasov11} . CEBs represent an appealing alternative to the
common Transition Edge Sensor technology.

In our design, the CEB absorber is capacitively coupled to the incoming radiation collected by Frequency Selective Surface (FSS)
unit cells antennas. The absorbed power is read out by Superconductor-Insulator-Normal (SIN) junctions, and at the same time,
thanks to a negative electrothermal feedback, the same junctions remove hot electrons improving the time constant and the noise
property of the detector itself. To date CEBs have not been flown on a stratospheric balloons yet, neither observed astrophysical
sources. Here we present the performance of a CEBs array sample, suitable to be integrated in the focal plane of the 350$\,$GHz
channel of the OLIMPO experiment. OLIMPO is a balloon-borne experiment aimed at observing the S-Z effect of hundreds of galaxy
clusters, obtaining, for the first time, spectroscopic observations by means of a differential Fourier Transformer
Spectrometer\cite{Schillaci14} . The balloon will be launched from Svalbard islands for a long duration (two weeks) stratospheric
flight. In Sec.\ref{sec:setup} we describe the CEBs sample array and its test setup. We have performed optical tests illuminating
the sample with chopped (77-300)$\,$K blackbody radiation and tested its sensitivity to X-ray photons using a microfocus X-ray
source; these measurements are presented and discussed in Sec.$\,$\ref{sec: optmeas} and in Sec.$\,$\ref{sec: xmeas}, respectively.
We conclude in Sec.$\,$\ref{sec:concl} summarizing our results in the view of future tests.

\section{EXPERIMENTAL SETUP}\label{sec:setup}
\begin{figure}
   \begin{center}
   \begin{tabular}{c}
   \includegraphics[height=7cm]{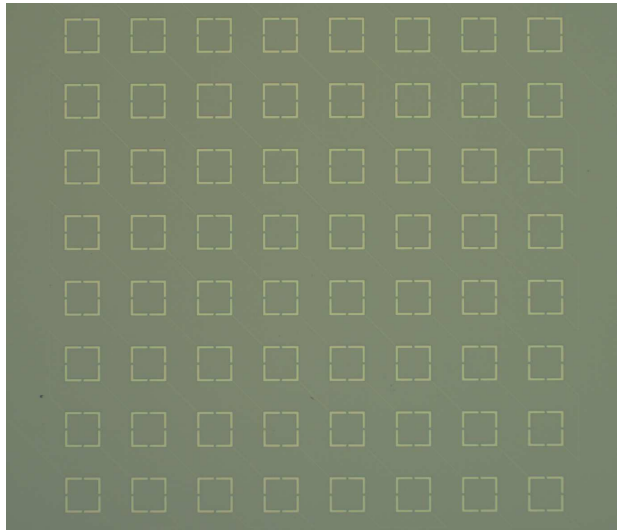}
   \end{tabular}
   \end{center}
   \caption[example]
   { \label{fig:sample}
Optical picture of the FSS array. The square structures represent the FSS element. Inside each FSS unit cell there is one square
element with 4 CEB detectors embedded in it. The side of each unit cell is 0.5$\,$mm.}
   \end{figure}

The CEBs sample array has been developed for the 350$\,$GHz channel of the OLIMPO experiment. The geometric properties of the
sample match the focal plane constraints and the optical properties of the telescope. The pixel of the tested sample is composed of
144 CEBs connected in series and in parallel (Fig.$\,$\ref{fig:sample}): this configuration sets the saturation limit of the sample
to about 41$\,$pW, which is the expected background load at balloon altitude in our 350$\,$GHz band for a diffraction-limited beam.
A HDPE window and a stack of metal mesh filters define the spectral bandwidth of the detectors. This, measured with a laboratory
FTS, results in a bandwidth 29$\%$ wide centered at 375$\,$GHz. The sample is optically coupled to the incoming radiation through a
couple of back-to-back horns with a cylindrical waveguide in between, and is integrated in a backshort cavity located at a distance
from the absorber which is $\lambda/4$ of the central wavelength. This optical configuration results from HFSS simulations which
have maximized the total energy collected. The sample is cooled down to 305$\,$mK with a two-stages He$^3$ fridge, pre-cooled by a
pulse tube refrigerator.

\begin{figure}
   \begin{center}
   \begin{tabular}{c}
   \includegraphics[height=7cm]{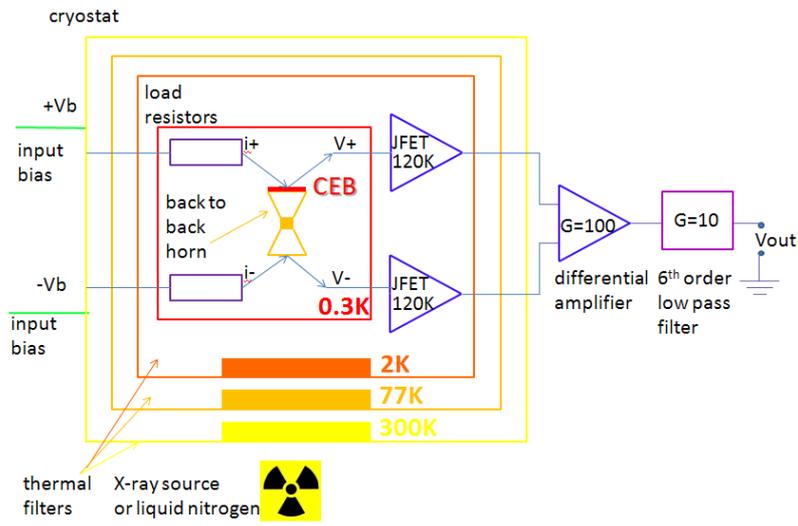}
   \end{tabular}
   \end{center}
   \caption[example]
   { \label{fig:testbed}
Block diagram of the experimental setup for optical measurements and irradiation of CEBs array with a microfocus X-ray source.}
   \end{figure}

A low noise readout electronics, on the 300$\,$K shield of the cryostat, and two 10$\,$MOhm load resistors very close to the sample
send the bias current to the CEBs array. A couple of JFET amplifiers, in the source-follower configuration already used for the HFI
instrument on board of the Planck satellite\cite{Brienza06} , mounted on the 2$\,$K flange of the cryostat and heated up to about
120$\,$K, preamplify the output voltage from the sample. A room temperature differential amplifier, with gain 100, reads the final
output voltage from the sample (Fig.$\,$\ref{fig:testbed}). Since in laboratory conditions the photonic background load is larger
than the saturation limit of the sample, we tried to decrease it in two different ways. We present the results of measurements
performed on two samples working in a bandwidth centered on 350$\,$GHz. In the first one the incoming background load was reduced
by a cold (0.3$\,$K) neutral density filter(aluminized polypropylene with spectral transmission about 0.5$\%$ in our working
bandwidth) and a cold black polyethylene filter (in-band transmission around 90$\%$); for the second sample we mounted a thin
copper disk, with a small (1.5 mm diameter) hole, at the horn aperture, which was originally 4.8$\,$mm in diameter.

\section{OPTICAL MEASUREMENTS}\label{sec: optmeas}
We present here the optical performance of the sample detector with the neutral density filter.

\begin{figure}
   \begin{center}
   \begin{tabular}{c}
   \includegraphics[height=7cm]{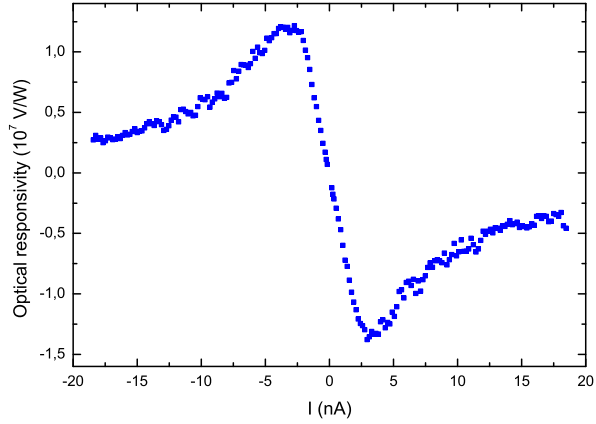}
   \end{tabular}
   \end{center}
   \caption[example]
   { \label{fig:optresp}
 Optical responsitivity versus bias voltage across the CEBs array.}
   \end{figure}

From the load curves performed illuminating the sample with a 300$\,$K and a 77$\,$K black body we estimated the voltage response
of the sample; at the optimal bias point this is about 0.47$\,$mV. To decrease the 50$\,$Hz pickup we performed all the optical
measurements with the pulse tube compressor temporarily off. Despite of the presence of Aluminum shields surrounding the high
impedance area (the sample and the two load resistors), the sample performance was limited by a disturbing pick-up at 100$\,$Hz.
The white noise level at the detector was about 190$\,$nV/sqrt(Hz). We measured the angular response of this system illuminating
the sample with a 350$\,$GHz high power parallel beam generated by a back wave oscillator. The beam pattern is approximately
Gaussian with a FWHM of about 36$^{\circ}$. With the knowledge of the spectral transmission of the filters stack and of the neutral
density filter we estimated the incident power load on the sample: when illuminated with chopped (300-77)$\,$K blackbody radiation
it resulted to be about 31$\,$pW. This estimate takes into account only the incoming background from the blackbody radiation. If
THz photons pass the quasioptical filters they can create quasiparticles in the CEB electrodes, modifying the load curve of the
device. These effects are difficult to model. However, on balloon experiment these signals will be reduced, so the lab test is a
stricter test for the performance of the detector. In our working conditions we estimated an optical responsitivity of about
$10^{7} V/W$ (Fig.$\,$\ref{fig:optresp}). We have illuminated the sample with a chopped (300-77)$\,$K blackbody radiation; this
resulted in an output voltage signal of about 100$\,\mu$Vpp at the optimal bias.


 We repeated the optical measurements using another sample with the copper disk in place of the neutral density filter.

The voltage response at the optimal bias point resulted to be about 1.7$\,$mV, despite of the presence of the copper. Further
reducing the incoming radiation flux with a smaller aperture in the copper disk would be challenging since, reducing the
millimetric radiation collected by the sample, it would degrade the signal to noise ratio. Illuminating the sample with chopped
(300-77)$\,$K blackbody radiation resulted in a voltage signal of 1.1$\,$mVpp at the detector. The presence of the copper disk on
the horn aperture changed the optical response: the effective angular response was approximately Gaussian with a FWHM of about
32$^{\circ}$. The incident power load was about 490$\,$pW, the optical responsitivity $R_{opt}$ of about 3.4$\,10^6 V/W$ and the
optical NEP of the order of 5.4$\,10^{-14}$W/sqrt(Hz). This value is significantly larger than the target value for the flight
detectors. The values currently achieved could be due a combination of intrinsic detector noise, readout noise, and imperfect
optical efficiency. More work is required to find the cause and improve. However, this performance is sufficient to study the
response to energetic radiation (see next paragraph).


\section{MEASUREMENT WITH A MICROFOCUS X-RAY SOURCE}\label{sec: xmeas}
\begin{figure}
   \begin{center}
   \begin{tabular}{c}
   \includegraphics[height=7cm]{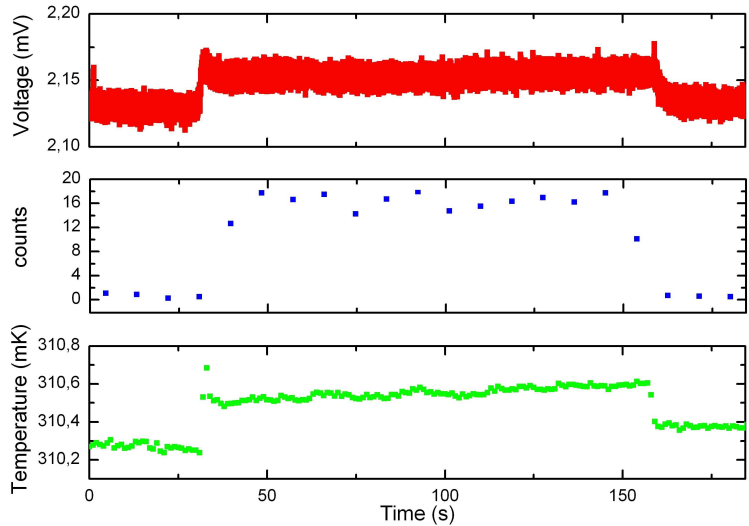}
   \end{tabular}
   \end{center}
   \caption[example]
   { \label{fig:offset}
Right: irradiation of a CEB sample with X-ray source. From top to bottom: output voltage from the detector, events recorded by a
Geiger counter 1$\,$m away from the source and evaporator temperature versus time.}
   \end{figure}

Using a x-ray source is the easiest way to perform a first test of sensitivity of incoherent detectors towards cosmic rays. As we
have already demonstrated\cite{Salatino14} , the energy deposited in our experiment is the same order of magnitude to the one that
a typical cosmic ray (100$\,$keV proton) should deposit interacting with a CEB. To the test the sensitivity of this sample to
cosmic rays we illuminated the sample with X-ray photons generated by a microfocus X-ray source (Hamamatsu model L10101) as already
done in a previous test\cite{Salatino14}. The CEB sample was biased optimally and the output voltage was filtered with a 6th order
low pass filter (200$\,$Hz cut-off, gain 10). The X-ray source can create large fluxes of photons with energy up to 100$\,$keV; we
operated it with different operating voltages and currents. The flux of high energy particles was monitored with a Geiger counter
near the cryostat. At high voltages of the tube ($>$60$\,$kV) both for low (1$\mu$A) and high (100$\mu$A) currents (low and high
flux of particle, respectively) we observed an increase of the temperature of the sample, and a positive offset in the output
voltage from the sample (Fig.$\,$\ref{fig:offset}). At high currents also the noise rms decreases during the irradiation. This is
expected since heating up the CEBs decreases their resistance, and this results, in turn, in a decrease of responsitivity. Here we
focus on the behavior with low currents, around 1$\mu$A.

\begin{figure}
   \begin{center}
   \begin{tabular}{c}
   \includegraphics[height=7cm]{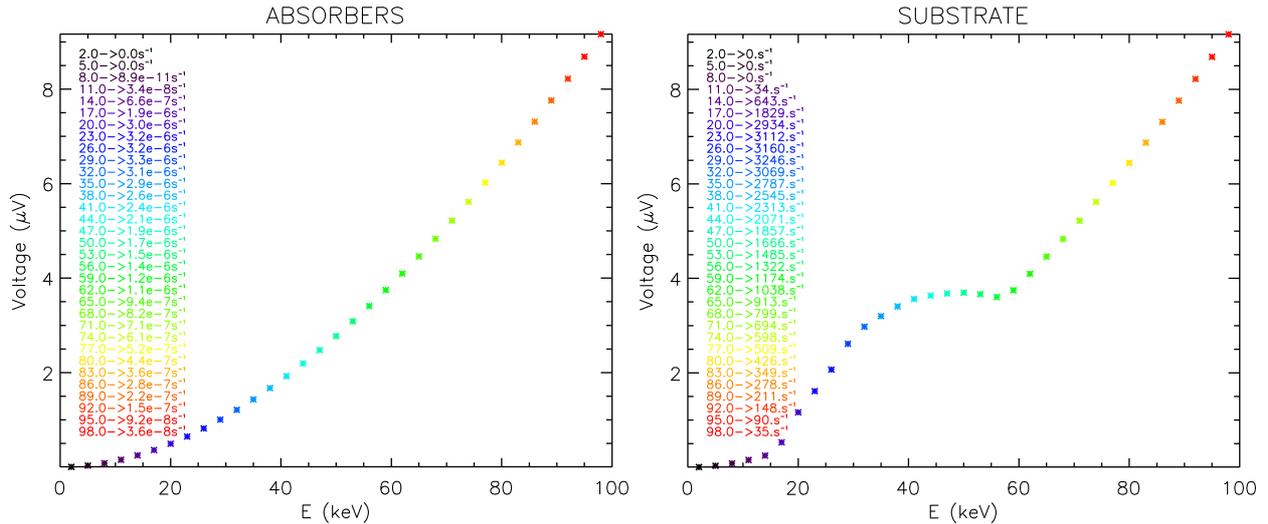}
   \end{tabular}
   \end{center}
   \caption[example]
   {\label{fig:DVavg}
Compton scattering between X-ray photons and the sample when the X-ray tube is operated at 100$\,$kV and 1$\mu$A. Left (right):
average voltage signals expected from interactions with absorbers (substrate) versus energy of the incoming photon. In the left
side the predicted event rate for each energy bin 3$\,$keV wide is reported.}
   \end{figure}

The spectrum of the X-ray photons produced by a microfocus X-ray source is described by the Kramers' law\cite{Klockenkämper97}. In
the absence of a calibrated detector, monitoring the flux of energetic photons emitted by the X-ray source, the proportionality
constant $K$ of the Kramers' law has been estimated in the following way:
    \begin{equation}
    \label{eq:K}
W \epsilon = K IZ  \int_{Emin}^{Emax}(E_{max}-E)dE \, ,
    \end{equation}
in eq.$\,$(\ref{eq:K}) $W=V\times I$ is the power produced by the X-ray tube operated with a voltage $V$ and a current $I$,
$\epsilon$ is the efficiency in the X-ray production. $\epsilon$ depends linearly on the operating voltage\cite{Compton63} and has
typical values of the order of 1$\%$. On the right-hand side we have the energy $E$ of the emitted X-ray photons weighted, with the
Kramers' law. The $E_{min}$ value is set by the Beryllium window of the source which removes the low-energy photons; the $E_{max}$
is set from the operating voltage of the tube itself. Operating the tube with different voltages and current results in a change of
the flux and of the energy of the X-ray photons. Typical values of the $K$ constant are of the order of $10^{28}$J$^{-1}$A$^{-1}$
and the number of the X-ray photons produced, for each energy bin 3$\,$keV wide, is in the range $(10^5\div10^{10})$/s. The flux
received at the detector is reduced by the absorption of the cryostat windows and filters stack, and by the ratio between the
target area and the beamwidth of the source (42$^{\circ}$ of aperture) projected at the distance of the sample (which in our case
was 17$\,$cm). We have studied the sensitivity of the CEBs sample considering two different cases with different sensitive areas.
In the first case the whole sample substrate (made of Silicon with thickness 0.28$\,$mm) is assumed to be sensitive to energetic
particles: however only an area as wide as the aperture of the cylindrical waveguide (0.9$\,$mm diameter) receives X-rays. In the
second case we assume that only the CEBs absorbers made of Aluminum inside the area of the cylindrical waveguide are sensitive to
the energetic particles: the total sensitive area in this case is 16$\,\mu$m$^2$, with a thickness of 10$\,$nm. The probability of
having an interaction in a target area $i$ is given by $\sigma_i/m_i \rho_i d_i$ where $\sigma_i/m_i$ is the Compton cross
section\cite{Xcom} , $\rho_i$ the density and $d_i$ the thickness. The expected rate is calculated as the resulting flux on each
target area times the corresponding interaction probability via Compton scattering. We can have less or more than one event within
the time constant $\tau$ of our electronics chain (i.e. 0.8$\,$ms). In the first case a voltage signal produced by the interaction
of a given energetic particle with the CEBs absorber is $R_{opt} \Delta E/\tau$ with $\Delta E$ the energy deposited by Compton
scattering. In the second case the signal can be estimated as $R_{opt} \Delta E R_{sub(abs)}$, with $R_{sub(abs)}$ the event rate
on the substrate (absorber). For the absorber the event rate is always smaller than the time constant, for the substrate it can be
larger or smaller depending on current an acceleration voltage in the source. We have binned the energy distribution of the emitted
X-ray photons in bins 3$\,$keV wide, and calculated, for the central energy of each bin, the expected signal produced by Compton
scattering. Then we have summed all the expected voltage signals weighting each of them with its corresponding probability of
interaction. In Fig.$\,$\ref{fig:DVavg} we report the mean voltage signals for a given operating condition of the X-ray tube
(1$\mu$A and 100$\,$keV) expected in the two cases: interactions on the CEBs absorber or interactions with the whole sample
substrate. At low currents, for some photon energy (for example E$<$20$\,$keV and $>$60$\,$keV in Fig.$\,$\ref{fig:DVavg}) the
expected signal produced by the interaction of a single photon with the absorber or the substrate is the same. During the
measurements with the X-ray source we recorded an increase in the temperature of the coldest flange of the $^3$He refrigerator.
Even for small temperature increases, around 1$\,$mK or above, the optical responsivity changed with respect to the value estimated
in Sec.$\,$\ref{sec: optmeas}. For simplicity, here we focus on cases with small temperature increases ($<1\,$mK), where we can
assume a linear behavior of the system, and avoid larger ones.


Modelling the interaction between an energetic particle with a CEB is complicated by the difficulty in separating the behavior of
the phononic system from the electronic one. Moreover, given the presence of the back-to-back horn the X-ray photons illuminates
only the substrate and the CEBs absorbers inside the area projected by the cylindrical waveguide, while the behavior of the
surrounding substrate and absorbers remains very difficult to model. Here we model the interaction considering only the Compton
scattering with the encountered material, disregarding the behavior of the area of the sample not reached by X-rays.

\begin{figure}
   \begin{center}
   \begin{tabular}{c}
   \includegraphics[height=7cm]{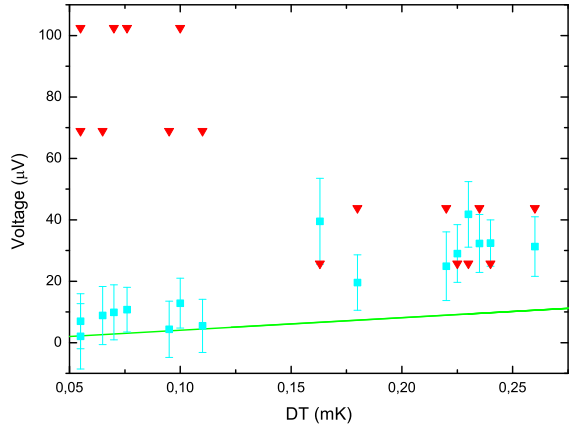}
   \end{tabular}
   \end{center}
   \caption[example]
   {\label{fig:DVtot}
Comparison between the measured output voltage offsets (squares) during X-ray irradiation, the voltage offset produced by thermal
heating of the evaporator flange (line) and the expected offset signal produced by interaction of X-ray photons with the sample
substrate in one time constant of the readout electronics (triangles). The X-ray tube was operated at 1$\,\mu$A in the voltage
range (70$\div$100)$\,$kV.}
   \end{figure}

We model the observed voltage offset as due to two contributions only: a thermal one produced by the heating of the evaporator
flange irradiated with X-ray photons, and one due to the Compton scattering between the sample (substrate and absorber) and X-ray
photons. We have calibrated the response of the evaporator to external heat dissipating power through a 10$\,$k$\Omega$ resistor
mounted on the 300$\,$mK stage and measuring the resulting temperature increase (Fig.$\,$\ref{fig:DVtot}). We compared the measured
offset voltage values with the estimated thermal contribution and the predicted voltage signal created by interactions between the
X-ray photons and the substrate. From the average voltage signal of Fig.$\,$\ref{fig:DVavg} we estimate the voltage offset produced
by events which happen during a single time constant. For interactions with the substrate this results to be between 43 and
102$\,\mu$V when the X-ray tube was operated between 70 and 100$\,$kV, respectively and at the same current 1$\,\mu$A. In the case
of interactions with the absorbers the expected values are negligible (of the order of $10^{-11}V$). We operated the X-ray tube in
these configurations for a total time of 32 minutes. Given our exposition time (and the sensitive area of the CEBs absorber, i.e.
16$\,\mu$m$^2$ irradiated with the X-ray source) we do not expect any event from the interaction between X-ray photons and the
absorber; on the contrary, we expect to be able to study the interaction between the X-ray photons and the substrate. In
Fig.$\,$\ref{fig:DVtot} is evident a mismatch between the observed voltage offset and the thermal contribution in the same
conditions. Our interpretation is that such a mismatch can be due to interaction with X-ray photons. However, the measured offset
is less than expected in case the interactions with the substrate are sensed by the detector. This means that the coupling between
the substrate and the absorbers is very weak, resulting in insensitivity of the detector to high energy particles hits. In our
working conditions the total number of X-ray photons which reaches the detector after transmission through our window and stack
filters is about 6.1 and 7.6$\,10^{10}$ particles per second. The cosmic rays flux produces energy depositions similar to the ones
used here, but the flux is many orders of magnitude lower than the flux of X rays we have used. This means that our detector is
suitable for operation in space, with evident advantages with respect to competing detectors like KIDs or TESs.



\section{Conclusions}\label{sec:concl}
We presented here the results concerning optical tests on two CEBs sample array, one of them suitable for integrating in the OLIMPO
focal plane at 350$\,$GHz. In this case the electrical responsivity was of the order of 10$^{7}V/W$. The optical tests we performed
have demonstrated the sensitivity of the CEBs towards the laboratory environment (i.e. 50$\,$Hz pickup, microphonics, and the
300$\,$K radiative background load). In particular the operation of the pulse tube compressor generates 1$\,$Hz vibrations  and 50
Hz pickup which reduce the CEBs performance. Performing a very short measurement, shutting down the compressor, was possible but it
is not totally reliable since the background load on the detector increases fast, given the increase of the far-infrared emission
of the shields surrounding the detector itself. In the correct load background, see Sec.$\,$\ref{sec: optmeas}, the sample
demonstrated a good optical responsivity, which needs to be optimized for future sub-mm and mm sky surveys. The noise level, on the
contrary, remained larger than the level required for an astronomical survey. This proved the necessity of developing a more
suitable testbed for these detectors, as a liquid helium cryostat, and more accurate shielding of the high-impedance area of the
testbed itself, and optimizing the optical coupling. We tested the response of such samples under high-energy particles
irradiation, improving the analysis with respect to previous work \cite{Salatino14} , in particular taking into account the energy
spectrum of the photons emitted by the X-ray tube. When the X-ray tube was operated with low current (1$\mu$A) and high voltage
(70-100$\,$kV) we observed a positive voltage offset at the output of our readout electronics that cannot be explained as due to
only to a thermal effect. However the level of the offset signal is low with respect to the signal produced by a large number of
interactions. This results seems to indicate (in our simple model) that only a few X-ray photons have interacted with the CEBs
detector, producing the mismatch voltage offset. These results need to be confirmed with an improved testbed with reduced
electromagnetic interference and readout and with a study of the interactions when the X-ray tube is operated at large currents.
Moreover, in order to test the sensitivity of the absorbers, we need to increase the total exposition time of our measurements, and
to remove the back-to-back horn in order to irradiate the entire CEBs absorbers area present in the sample.

\acknowledgments The authors wish to thank Dr. A. Schillaci and G. D'Alessandro for the spectral efficiency measurements, Dr. P.
Benedetti for manufacturing the pcb for the CEBs sample array. Moreover, we thank Dr. D. Fargion and Dr. I. Dafinei for allowing us
to use some of their instrumentation for our measurements.

\bibliography{biblist}   

\begin{thebibliography}{}

\bibitem{Planck13} P.A.R. Ade and N. Aghanim and C. Armitage-Caplan and M. Arnaud and M. Ashdown and F. Atrio-Barandela
    and J.Aumont and C. Baccigalupi and A.J. Banday and R.B. Barreiro and J.G. Bartlett and E. Battaner and K. Benabed and A. Benoît
    and A. Benoit-Lévy and J.-Ph. Bernard and M. Bersanelli and P. Bielewicz and J. Bobin and J.J. Bock and A. Bonaldi and J.R. Bond
    and J. Borrill and F.R. Bouchet and M. Bridges and M. Bucher and C. Burigana and R.C. Butler and E. Calabrese and B. Cappellini
    and J.-F. Cardoso and A. Catalano and A. Challinor and A. Chamballu and R.-R. Chary and X. Chen and H.C. Chiang and L.-Y. Chiang and
    and P.R. Christensen and S. Church and D.L. Clements and S. Colombi and L.P.L. Colombo and F. Couchot and A. Coulais and B.P. Crill et al.
    ``Planck 2013 results. XVI. Cosmological parameters", astro-ph/1303.5076 (2013).

\bibitem{BICEP2} P.A.R. Ade and R.W. Aikin and D. Barkats and S.J. Benton and C.A. Bischoff and J.J. Bock
    and J.A. Brevik and I. Buder and E. Bullock and C.D. Dowell and L. Duband and J.P. Filippini and S. Fliescher and S.R. Golwala
    and M. Halpern and M. Hasselfield and S.R. Hildebrandt and G.C. Hilton and V.V. Hristov and K.D. Irwin and K.S. Karkare and
    J.P. Kaufman and B.G. Keating and S.A. Kernasovskiy and J.M. Kovac and C.L. Kuo and E.M. Leitch and M. Lueker and P. Mason and
    C.B. Netterfield and H.T. Nguyen and R. O'Brient and R.W. Ogburn and A. Orlando and C. Pryke and C.D. Reintsema and S. Richter and
    R. Schwarz and C.D. Sheehy and Z.K. Staniszewski and R.V. Sudiwala and G.P. Teply and J.E. Tolan and A.D. Turner and A.G. Vieregg
    and C.L. Wong and K.W. ``Detection of B-mode polarization at degree angular scales", astro-ph/1403.3985 (2014).

\bibitem{Planck13b} P.A.R. Ade and N. Aghanim and C. Armitage-Caplan and M. Arnaud
    and M. Ashdown and F. Atrio-Barandela and J. Aumont and C. Baccigalupi and A.J. Banday and R.B. Barreiro and E. Battaner
    and K. Benabed and A. Benoît and A.Benoit-Lévy and J.-Ph. Bernard and M. Bersanelli and P. Bielewicz and J. Bobin and J.J. Bock
    and J.R. Bond and J. Borrill and F.R. Bouchet and M. Bridges and M.Bucher and C. Burigana and J.-F. Cardoso and A. Catalano and A.
    Challinor and A. Chamballu and L.-Y. Chiang and H.C. Chiang and P.R. Christensen and S. Church and D.L. Clements and S. Colombi
    and L.P.L. Colombo and F. Couchot and A. Coulais and B.P. Crill and A. Curto and F. Cuttaia and L. Danese et al. ``Planck 2013 results X. 
    Energetic particle effects: characterization, removal, and simulation", astro-ph/1303.5071 (2013).

\bibitem{Catalano14} A. Catalano and P.A.R. Ade and Y. Atik and A. Benoit and E. Bréele and J.J. Bock and P. Camus and M. Chabot and M.Charra
    and B.P. Crill and N. Coron and A. Coulais and F.-X. Désert and L. Fauvet and Y. Giraud-Héraud and O. Guillaudin and W. Holmes
    and W.C. Jones and J.-M. Lamarre and J.Macías-Pérez and M. Martinez and A. Miniussi and A.Monfardini and F. Pajot and G. Patanchon
    and A. Pelissier and M. Piat and J.-L. Puget and C. Renault and C. Rosset and D. Santos and A. Sauvé and L.D. Spencer and R. 
    ``Impact of particles on the Planck HFI detectors: Ground-based measurements and physical interpretation", astro-ph/1403.6592 (2014).

\bibitem{BICEP3} P.A.R. Ade and R.W. Aikin and M. Amiri and D. Barkats and S.J. Benton and C.A. Bischoff and J.J. Bock
    and J.A. Brevik and I. Buder and E. Bullock and G. Davis and C.D. Dowell and L. Duband and J.P. Filippini and S. Fliescher and S.R. Golwala
    and M. Halpern and M.Hasselfield. ``BICEP2 II: Experiment and Three-Year Data Set", astro-ph/1403.4302 (2014).

\bibitem{SWIPE12} P. de Bernardis and S. Aiola and G. Amico and E. Battistelli and A. Coppolecchia and A. Cruciani and A. D'Addabbo and G. D'Alessandro
    and S. De Gregori and M. De Petris and D. Goldie and R. Gualtieri and V. Haynes and L. Lamagna and B. Maffei and S. Masi and F. Nati and W. M. Ng
    and L. Pagano and F. Piacentini and L. Piccirillo and G. Pisano and G. Romeo and M. Salatino and A. Schillaci and E. Tommasi and S. Withington.
    ``SWIPE: a bolometric polarimeter for the Large-Scale Polarization Explorer", Proceedings of the SPIE, astro-ph/1208.0282", vol. 8452, article id. 84523F, 
    astro-ph/1208.0282 (2012).

\bibitem{Fraisse13} A.A. Fraisse and P.A.R. Ade and M. Amiri and S.J. Benton and J.J. Bock and J.R. Bond and J.A. Bonetti and S. Bryan and B. Burger
    and H.C. Chiang and C.N. Clark and C.R. Contaldi and B.P. Crill and G. Davis and O. Doré and M. Farhang and J.P. Filippini and L.M. Fissel
    and N.N. Gandilo and S. Golwala J.E. Gudmundsson and M. Hasselfield and G. Hilton and W. Holmes and V.V. Hristov and K. Irwin and W.C. Jones
    and C.L. Kuo and C.J. MacTavish and P.V. Mason and T.E. Montroy and T.A. Morford and C.B. Netterfield and D.T. O'Dea and A.S. Rahlin
    and C. Reintsema and J.E. Ruhl and M.C. Runyan and M. A. Schenker and J.A. Shariff and J.D. Soler and A. Trangsrud and C. Tucker
    and R.S. Tucker and A.D. Turner and D. Wiebe, ``SPIDER: probing the early Universe with a suborbital polarimeter", Journal of Cosmology and Astroparticle Physics,
    issue 04, article id. 047 (2013).

\bibitem{PRISM14} P. Andr$\acute{e}$ and C. Baccigalupi and A. Banday and D. Barbosa and B. Barreiro and J. Bartlett and N. Bartolo and E. Battistelli and R. Battye
              and G. Bendo and A. Benoit and J.-Ph. Bernard and M. Bersanelli and M. Béthermin and P. Bielewicz and A. Bonaldi and F. Bouchet
              and F. Boulanger and J. Brand and M. Bucher and C. Burigana and Z.-Y. Cai and P. Camus and F. Casas and V. Casasola and G. Caste
              and A. Challinor and J. Chluba and G. Chon and S. Colafrancesco and B. Comis and F. Cuttaia and G. D'Alessandro and
              A. Da Silva and R. Davis and M. de Avillez and P. de Bernardis and M. De Petris and A. de Rosa and G. de Zotti and J. Delabrouille
              and F.-X. Désert and C. Dickinson and J.M. Diego and J. Dunkley and T. Enßlin and J. Errard and E. Falgarone and P. Ferreira
              and K. Ferrière and F. Finelli Fabio and A. Fletcher and P. Fosalba and G. Fuller and S. Galli, Silvia and K. Ganga and J. García-Bellido
              and A. Ghribi and M. Giard and Y. Giraud-Héraud and J. Gonzalez-Nuevo and K. Grainge and A. Gruppuso and A. Hall
              and J.-Ch. Hamilton and M. Haverkorn and C. Hernandez-Monteagudo and D. Herranz and M. Jackson and A. Jaffe and R.
              Khatri and M. Kunz and L. Lamagna and M. Lattanzi and P. Leahy and J. Lesgourgues and M. Liguori and E. Liuzzo
              and M. Lopez-Caniego and J. Macias-Perez and B. Maffei and D. Maino and A. Mangilli and E. Martinez-Gonzalez and C.J.A.P.
              Martins and S. Masi and M. Massardi and S. Matarrese and A. Melchiorri and J.-B. Melin and A.
              Mennella and A. Arturo and M.-A. Miville-Deschênes and A. Monfardini and A. Murphy and P. Naselsky
              and F. Nati and P. Natoli and M. Negrello and F. Noviello and C. O'Sullivan and F. Paci and L. Pagano and
              R. Paladino and N. Palanque-Delabrouille and D. Paoletti and H. Peiris and F. Perrotta and F. Piacentini
              and M. Piat and L. Piccirillo and G. Pisano and G. Polenta and A. Pollo and N. Ponthieu and
              M. Remazeilles and S. Ricciardi and M. Roman and C. Rosset and J.-A. Rubino-Martin and M. Salatino
              and A. Schillaci and P. Shellard and J. Silk and A. Starobinsky and R. Stompor and R. Sunyaev and A. Tartari
              and L. Terenzi and L. Toffolatti and M. Tomasi and N. Trappe and M. Tristram and T. Trombetti and
              M. Tucci and R. Van de Weijgaert and B. Van Tent and L. Verde and B. Vielva and B. Wandelt and R. Watson and S.
              Withington, ``PRISM (Polarized Radiation Imaging and Spectroscopy Mission): an extended white paper", Journal of Cosmology and Astroparticle Physics,
              issue 02, article id. 006 (2014).

    \bibitem{Golubev01} D. Golubev and L.S. Kuzmin,
    ``Nonequilibrium theory of a hot-electron bolometer with normal metal-insulator-superconductor tunnel junction", Journal of Applied Physics,
    vol. 89, issue 11, pag. 6464-6472 (2001).

    \bibitem{Kuzmin02} L.S. Kuzmin,
    ``Optimization of the Hot-Electron Bolometer and A Cascade Quasiparticle Amplifier for Space Astronomy",
    International Workshop on Superconducting Nano-Electronics Devices, vol. 89, issue 11, pag. 145-154 (2002).

    \bibitem{Kuzminb02} L.S. Kuzmin and D. Golubev,
    ``On the concept of an optimal hot-electron bolometer with NIS tunnel junctions", Physica C,
    vol. 372-376, pag. 378-382 (2002).

    \bibitem{Tarasov11} M.A. Tarasov and L.S. Kuzmin and V.S. Edelman and S. Mahashabde and P. de Bernardis,
    ``Optical Response of a Cold-Electron Bolometer Array Integrated in a 345-GHz Cross-Slot Antenna", IEEE Trans. Appl. Superconduct.,
    vol. 21, no. 6, pag. 3635 – 3639 (2011).

    \bibitem{Schillaci14} A. Schillaci and G. D'Alessandro and P. de Bernardis and S. Masi and C. Paiva Novaes and M. Gervasi and M. Zannoni. ``
Efficient Differential Fourier-Transform Spectrometer for precision Sunyaev-Zel'dovich effect measurements", Astronomy and Astrophysics, vol. 565, id. A125, astro-ph/1402.4091 
    (2014).

    \bibitem{Brienza06} D. Brienza and L. De Angelis and P. de Bernardis and F. Era and M.C. Falvella and S. Malatesti and S. Masi and F. Nati and L. Nati and F. Piacentini and G. Polenta. ``Cryogenic Preamplifiers for high resistance bolometers". Seventh International Workshop on Low Temperature Electronics, WOLTE-7 - ESA-WPP-264,
    pag. 283-288 (2006).

    \bibitem{Salatino14} M. Salatino and P. de Bernardis and L.S. Kuzmin and S. Mahashabde and S. Masi. ``Sensitivity to Cosmic Rays of Cold Electron Bolometers for Space Applications". Journal of Low Temperature Physics, vol. 176 issue 3, pag. 323-330, 10.1007/s10909-013-1057-5 (2014).

    \bibitem{Klockenkämper97} R. Klockenk$\ddot{a}$mper, ``Total-reflection X-ray fluorescence analysis", John Wiley and Sons, New York, pag. 13 (1997).

    \bibitem{Compton63} A.H. Compton and S.K. Allison. ``X-Rays in Theory and Experiment". D. Van Nostrand Company, Inc., Princeton, pag. 89-90 (1963).

    \bibitem{Xcom} database NIST Xcom, http://physics.nist.gov/PhysRefData/Xcom/html/xcom1.html.

\end{thebibliography}
\bibliographystyle{spiebib}   

\end{document}